\documentclass[10.5pt,conference,compsocconf]{IEEEtran}
\usepackage{times,amsfonts,amsmath,graphicx,algorithm,algorithmic,caption,balance}
\usepackage[font=small]{caption}

\begin{document}
\pagenumbering{gobble}
%
\title{\textbf{\Large A Review of Audio Features and Statistical Models Exploited for Voice Pattern Design\\[0.2ex]}}

\author{\IEEEauthorblockN{~\\[-0.4ex]\large Ngoc Q. K. Duong\\[0.3ex]\normalsize}
\IEEEauthorblockA{Technicolor\\
975 avenue des {C}hamps {B}lancs\\
35576 Cesson S\'evign\'e, France\\
Email: {\tt quang-khanh-ngoc.duong@technicolor.com}}
\and
\IEEEauthorblockN{~\\[-0.4ex]\large Hien-Thanh Duong\\[0.3ex]\normalsize}
\IEEEauthorblockA{Faculty of Information Technology\\
Hanoi University of Mining and Geology\\
Hanoi city, Vietnam\\
Email: {\tt duongthihienthanh@humg.edu.vn}}
}
\maketitle

\begin{abstract}
Audio fingerprinting, also named as audio hashing, has been well-known as a powerful technique to perform audio identification and synchronization. It basically involves two major steps: fingerprint (voice pattern) design and matching search. While the first step concerns the derivation of a robust and compact audio signature, the second step usually requires knowledge about database and quick-search algorithms. Though this technique offers a wide range of real-world applications, to the best of the authors' knowledge, a comprehensive survey of existing algorithms appeared more than eight years ago. Thus, in this paper, we present a more up-to-date review and, for emphasizing on the audio signal processing aspect, we focus our state-of-the-art survey on the fingerprint design step for which various audio features and their tractable statistical models are discussed.
\end{abstract}


\IEEEpeerreviewmaketitle

\section{Introduction}
\label{sec:introduction}

Real-time user interactive applications have emerged nowadays thanks to the increased power of mobile devices and their Internet access speed. Let us consider applications like music recognition \cite{Wang-ISMIR03}\cite{Haitsma-ISMIR02}, e.g., people hear a song in a public place and they want to know more about it, or personalized TV entertainment \cite{Fink-ITV}\cite{WP2-ICCE11}, e.g., people want to see more service and related content on the Web in addition to the main view from TV; both require a fast and reliable audio identification system in order to match the observed audio signal with its origin stored in a large database. For these purposes, several research directions have been studied, such as audio fingerprinting \cite{Cano-WMSP02}, audio watermarking \cite{Kim-watermark}, and timeline insertion \cite{WP2-ICCE11}. While watermarking and timeline approaches both require to embed signature into the original media content, which is sometimes inconvenient for the considered applications, fingerprinting technique allows directly monitoring the data for identification. Hence, audio fingerprinting has been widely investigated in the literature and already been deployed in many commercialized products \cite{Wang-ISMIR03}\cite{Macrae-ICME10}\cite{TVplus}\cite{intonow}\cite{mediasync}\cite{Civolution}. This technique has recently been exploited for other applications such as media content synchronization \cite{Duong-ICCE2012}\cite{Duong-ICASSP2013}, multiple video clustering \cite{Ellis-ICASSP10},  repeating object detection \cite{Rafii-ICASSP14}, and live version identification \cite{Rafii-ICASSP14}.

\begin{figure}[htb]
\centering
\captionsetup{justification=centering}
\includegraphics[scale=0.65]{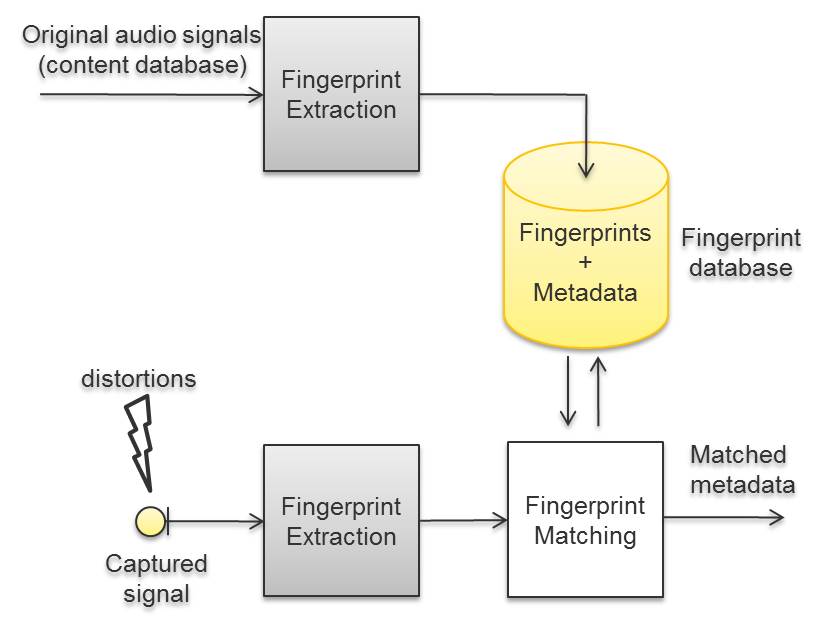}
\caption{General architecture of an audio fingerprinting system.} \label{fig:FPframework}
\end{figure}

A general architecture for an audio fingerprinting system, which can be used for either audio identification or audio synchronization purpose, is depicted in Fig. \ref{fig:FPframework}. The fingerprint extraction derives a set of relevant audio features followed by an optional post-processing and feature modeling. Fingerprints of the original audio collection and its corresponding metadata (e.g., audio ID, name, time frame index, etc.) are systematically stored in a database. Then given a short recording from the user side, its feature vectors (i.e fingerprints) are computed in the same way as they were for the original data. Finally, a searching algorithm will find the best match between these fingerprints with those stored in the database so that the recorded audio signal is labeled by the matched metadata.

In real-world recording, the audio signal often undergoes many kinds of distortion: acoustical reverberation, background noise addition, quantization error, etc. Thus, the derived fingerprints must be robust with respect to these various signal degradations. Beside, the fingerprint size should be as small as possible to save memory resources and to allow real-time matching. The details of general properties of the audio fingerprint was well-discussed in \cite{Wang-ISMIR03}\cite{Cano-APreview}\cite{Gael-IEEE2013}.
In order to fulfil those requirements, audio sample signal is often transformed into Time-Frequency (T-F) domain via the Short Time Fourier transform (STFT) \cite{Cano-APreview} where numerous distinguishable characteristics such as high-level musical attributes, e.g., predominant  pitch, harmony structure, or low level spectral features, e.g., mel-frequency cepstrum, spectral centroids, spectral note onsets, etc., are exploited. To further compact the fingerprints, some approaches continue to fit the spectral feature vectors to a statistical model, e.g., Gaussian Mixture Model (GMM) \cite{Ramalingam-IEEETransIFS}, Hidden Markov Model (HMM) \cite{Cano-CME04}, so that in the end only the set of model parameters are used as fingerprints.

Though diverse fingerprinting algorithms have been proposed in the literature, the number of review papers remains limited where, to the best of the authors' knowledge, a comprehensive review of fingerprinting algorithms was presented more than eight years ago \cite{Cano-WMSP02}\cite{Cano-APreview}, and a more recent survey \cite{Vijay-ISMIR11} only focusing on computer vision based approaches (e.g., methods proposed in \cite{Ke-CVPR05}\cite{Baluja-ICASSP07}). In this paper, we present a more up-to-date review of the domain, with particular focus concerning the fingerprint extraction block in Fig. \ref{fig:FPframework}, where various audio spectral features and their statistical models are summarized systematically. The presentation would particularly benefit new researchers in the domain and engineers in the sense that they would easily follow the described steps to implement different audio fingerprints.

The structure of the rest of the paper is as follows. We first present a general architecture for fingerprint design in Section \ref{sec:FPextraction}, we then review various audio features, which have been extensively exploited in the literature, in Section \ref{sec:features}. The detail of some statistical feature models is introduced in Section \ref{sec:models}. Finally, we conclude in Section \ref{sec:conclusion}.


\section{General architecture of fingerprint design}
\label{sec:FPextraction}

Fig. \ref{fig:FPextraction} depicts a general workflow of the fingerprint design. The purpose of each block is summarized as follows:

\begin{figure}[htb]
\centering
\captionsetup{justification=centering}
\includegraphics[scale=0.65]{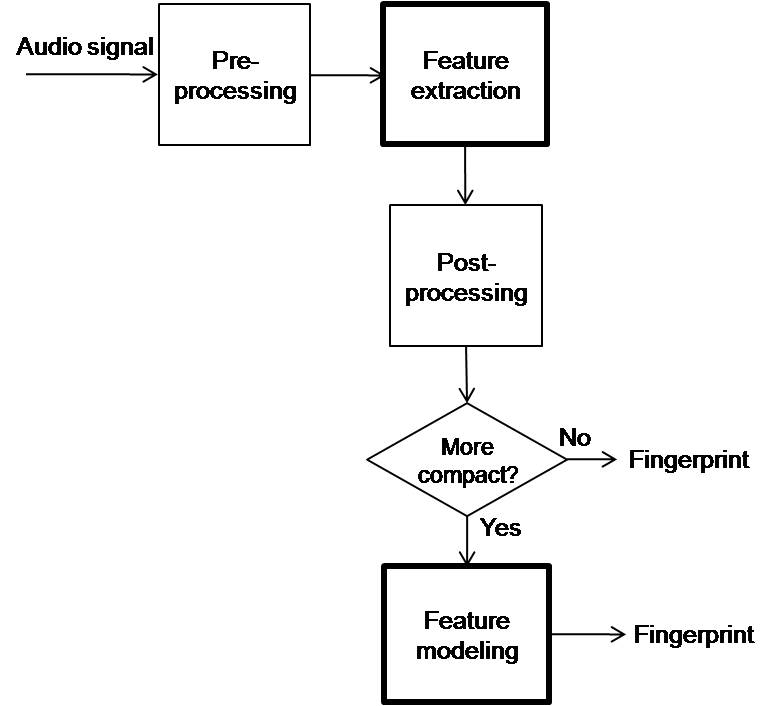}
\caption{General workflow of the fingerprint design.} \label{fig:FPextraction}
\end{figure}

\begin{itemize}
\item \emph{Pre-processing}: in this step, input audio signal is often first digitalized (if necessary), re-sampled to a target sampling rate, and bandpass filtered. Other types of processing includes decorrelation and amplitude normalization \cite{Cano-APreview}. Then the processed signal is segmented into overlapping time frames where a linear transformation, e.g., Fast Fourier Transform (FFT), Discrete Cosine transform (DCT), or wavelet transform \cite{Cano-APreview}, is applied to each frame. At this stage, the input time-domain signal is represented in a feature domain, and the most popular feature domain is time-frequency representation given by the STFT.
\item \emph{Feature extraction}: this is a major process since the choice of "which feature is used" will directly affect the performance of the entire fingerprinting system. A great diversity of features have been investigated targeting the reduction of dimensionality as well as the invariance to various distortions. For summary, most approaches first map the linear time-frequency representation given by the STFT to an auditory-motivated frequency scale, i.e., \ Mel, Bark, Log, or Cent scale, via filterbanks \cite{Haitsma-ISMIR02}\cite{Ramona-DAFX11}. This mapping step greatly reduces the spectrogram size since the number of filterbanks is usually much smaller than the FFT length. Then a feature vector such as Mel-Frequency Cepstral Coefficients (MFCC), spectral centroids of all subbands, etc., are computed for each time frame. In some systems, the first and second derivatives of the feature vectors are also integrated to better track the temporal variation of audio signals \cite{Cano-APreview}\cite{Ramalingam-IEEETransIFS}. Other types of feature that worth mentioning are e.g.,\  time localized frequency peak \cite{Richard-ICASSP2010}, time-frequency energy peak location \cite{Wang-ISMIR03}, or those developed in image processing based approaches such as top-wavelet coefficients computed on the spectral image \cite{Baluja-ICASSP07} and multiscale Gabor atoms extracted by Matching Pursuit algorithm \cite{Ellis-ICASSP10}. Recently, a general framework for dictionary based feature learning has also been introduced \cite{Daudet-ICASSP14}.
\item \emph{Post-processing}: the feature vectors computed in the previous step are often real-valued and the absolute range depends on the signal power. Therefore when Euclidean distance is used in the matching step, mean substraction and component wise variance normalization are recommended \cite{Seo-ICASSP05}\cite{Schedl-MIREX11}. Another popular post-processing is quantization where each entry of the feature vectors is quantized to a binary number in order to gain robustness against distortions and, more importantly, to obtain memory efficiency \cite{Haitsma-ISMIR02}\cite{BestpaperICME2012}\cite{Baluja-ICASSP07}
    \cite{Rafii-ICASSP14}. In many existing system, fingerprint is achieved after this step.
\item \emph{Feature modeling}: this block is sometimes deployed in order to further compact the fingerprint. In this case, a large number of feature vectors along time frames is fitted to a statistical model so that an input audio signal is well-characterized by the model parameters, which are then stored as a fingerprint \cite{Cano-AES02}\cite{Ramalingam-IEEETransIFS}\cite{Hao-IEEETransASLP10}. Popular model includes Gaussian Mixture Model (GMM), Hidden Markov Model (HMM). Other approaches used decomposition techniques, e.g., Non-negative MAtrix Factorization (NMF), to help decreasing data dimension and therefore to reduce the local statistical redundancy of the feature vectors \cite{Burges-IEEETransASLP}\cite{Deng-ICCT11}.
\end{itemize}

Since the pre-processing and post-processing steps are quite straightforward, in the following of the paper we will present more detail only on the feature extraction and the feature modeling blocks.

\section{Feature extraction}
\label{sec:features}

Summarizing numerous types of audio features used for the fingerprint design so far will certainly go beyond the scope of this paper. Thus in this section, we select to present the most popular low level features in the spectral domain only.

\subsection{MFCC}
\label{ssec:mfcc}

MFCC is one of the most popular feature considered in speech recognition where the amplitude spectrum of input audio signal is first weighted by triangular filters spaced according to the Mel scale, and DCT is then applied to decorrelate the Mel-spectral vectors. MFCC was shown to be applicable for music signal also in \cite{Logan-ISMIR00}. Examples of fingerprinting algorithms used MFCC feature are found in \cite{Logan-ISMIR00}\cite{Ramalingam-IEEETransIFS}. In \cite{Ashish-EUSIPCO13}, MFCC was used also for clustering and synchronizing large scale audio-video sequences recorded by multiple users during an event. Matlab implementations for the computation of MFCC are available \cite{footnote1}\cite{footnote2}. 

\subsection{Spectral Energy Peak (SEP)}
\label{ssec:sep}

SEP for music identification systems was described in \cite{Yang-WASPAA01}\cite{Wang-ISMIR03} where a time-frequency point is considered as a peak if it has higher amplitude than its neighboring points. SEP is argued to be intrinsically robust to even high level background noise and can provide discrimination in sound mixtures \cite{Ellis-ICASSP07}. In well-known Shazam's system \cite{Wang-ISMIR03} time-frequency coordinates of the energy peaks was described as sparse landmark points. Then by using pairs of landmark points rather than single points, the fingerprints exploited the spectral structure of sound sources. This landmark feature can also be found in \cite{Ellis-ICASSP10} and \cite{Bryan-ICASSP12} for multiple video clustering. Ramona \emph{et al.} used start times of the spectral energy peaks, referred to as onsets, for the automatic alignment of audio occurrences in their fingerprinting system \cite{Ramona-DAFX11}\cite{Ramona-ICASSP11}.

\subsection{Spectral Band Energy (SBE)}
\label{ssec:sbe}

Together with spectral peak, SBE has been widely exploited in fingerprinting algorithms. Let us denote by $s(n,f)$ a STFT coefficient of an audio signal at time frame index $n$ and frequency bin index $f$, $1\leq f \leq M$. Let us also denote by $b$ an auditory-motivated subband index, i.e., in either Mel, Bark, Log, or Cent scale, and $l_b$ and $h_b$ the lower and upper edges of $b$-th subband. SBE is then computed, with normalization, in each time frame and each frequency subband range by
\begin{equation}
F_{n,b}^{\text{SBE}}=\frac{\sum_{f=l_b}^{h_b}|s(n,f)|^2}{\sum_{f=1}^M|s(n,f)|^2}.
\label{eq:sbe}
\end{equation}

Haitsma et al. proposed a famous fingerprint in \cite{Haitsma-ISMIR02} where SBEs were first computed in a block containing 257 time frames and 33 Bark-scale frequency subbands, then each $F_{n,b}^{\text{SBE}}$ was quantized to a binary value (either 0 or 1) based on its differences compared to neighboring points. Other fingerprinting algorithms exploiting SBE feature were found for instance in \cite{Allamanche-ISMIR02}\cite{Ramalingam-IEEETransIFS}. Variances of this subband energy difference feature can be found in more recent approaches \cite{BestpaperICME2012}\cite{Ke-CVPR05}.

\subsection{Spectral Flatness Measure (SFM)}
\label{ssec:sfm}

SFM, also known as Wiener entropy, relates to the tonality aspect of audio signals and it is therefore often used to distinguish different recordings. SFM is computed in each time-frequency subband point $(n,b)$ as
\begin{equation}
F_{n,b}^{\text{SFM}}=\frac{\big(\prod_{f=l_b}^{h_b}|s(n,f)|^2\big)^{\frac{1}{h_b-l_b+1}}}{\frac{1}{h_b-l_b+1}\sum_{f=l_b}^{h_b}|s(n,f)|^2}.
\label{eq:sfm}
\end{equation}

A high SFM indicates the similarity of signal power over all frequencies while a low SFM means that signal power is concentrated in a relatively small number of frequencies over the full subband.

A similarly feature to SFM, which is also a measure of the tonal-like or noise-like characteristic of audio signal and was exploited as fingerprint, is spectral crest factor (SCF). SCF is computed by
\begin{equation}
F_{n,b}^{\text{SCF}}=\frac{\text{max}_{f\in[l_b,h_b]}\big(|s(n,f)|^2\big)}{\frac{1}{h_b-l_b+1}\sum_{f=l_b}^{h_b}|s(n,f)|^2}.
\label{eq:scf}
\end{equation}

SFM and SCF were found to be the most promising features for audio matching with common distortions in \cite{Herre-WASPAA01} and were both considered in other fingerprinting algorithms \cite{Allamanche-ISMIR02}\cite{Ramalingam-IEEETransIFS}.

\subsection{Spectral Centroid (SC)}
\label{ssec:sc}

SC is also a popular measure used in audio signal processing to indicate where the "center of mass" of a subband spectrum is. It is formulated as
\begin{equation}
F_{n,b}^{\text{SC}}=\frac{\sum_{f=l_b}^{h_b}f.|s(n,f)|^2}{\sum_{f=l_b}^{h_b}|s(n,f)|^2}.
\label{eq:sc}
\end{equation}

SC was argued to be robust over equalization, compression, and noise addition. It was reported in \cite{Seo-ICASSP05} and \cite{Ramalingam-IEEETransIFS} that SC-based fingerprints offered better audio recognition than MFCC-based fingerprints with 3 to 4 second length audio clips. In our preliminary experiment with speech utterances distorted by reverberation and real-world background noise, we also observed that SC-based fingerprints resulted in higher recognition accuracy than MFCC-, SBR-, and SFM-based fingerprints without post-processing.

Given one of the feature parameters $F_{n,b}$ computed in each time-frequency subband point $(n,b)$ as described above, a $d$-dimensional feature vector $\mathbf{F}_n=[F_{n,1},...,F_{n,d}]^T$ is formed to describe the corresponding characteristic of the signal at time frame $n$, where $T$ denotes vector transpose and $d$ is the total number of subbands. When the first and second derivatives of the feature vectors are additionally considered, for better characterizing the temporal variation of audio signal, $\mathbf{F}_n$ will then be a $3d$-dimensional vector \cite{Ramalingam-IEEETransIFS} before passing to the post-processing block shown in Figure \ref{fig:FPextraction}.

\section{Feature modeling}
\label{sec:models}

In some systems, in order to further compact the fingerprint the feature vectors $\mathbf{F}_n$ can be adapted to a statistical model. This step allows to reduce the global redundancy of spectral features. As a result, a long sequence of feature vectors $\mathbf{F}_n, n=1,...,N$ is characterized by a significantly smaller number of the model parameters while ensuring the discriminative power. In this section we review the use of three popular models, namely gaussian mixture model (GMM), hidden Markov model (HMM), and nonnegative matrix factorization (NMF), for the fingerprint design.

\subsection{GMM-based fingerprint}
\label{ssec:GMM}

GMM has been used to model the spectral shape of audio signals in many different applications ranging from speaker identification \cite{Reynold-ASLP95} to speech enhancement \cite{Hao-IEEETransASLP10}, etc. It was also investigated for audio fingerprinting by Ramalingam and Krishnan \cite{Ramalingam-IEEETransIFS}, where spectral feature vectors $\mathbf{F}_n$ are modeled as a multidimensional K-state Gaussian mixture with probability density function (pdf) given by
\begin{equation}
p(\mathbf{F}_n)=\sum_{k=1}^{K}\alpha_{k}\mathcal{N}_c(\mathbf{F}_n|\mathbf{\mu}_k,\mathbf{\Sigma}_k)
\label{eq:GMMpdf}
\end{equation}
where $\alpha_k$, which satisfies $\sum_{k=1}^K \alpha_k=1$, $\mathbf{\mu}_k$ and
$\mathbf{\Sigma}_k$ are the weight, the mean vector and the covariance matrix of the $k$-th state, respectively, and
\begin{equation}
\mathcal{N}_c(\mathbf{F}_n|\mathbf{\mu}_k,\mathbf{\Sigma}_k)=\frac{1}{|\pi
\mathbf{\Sigma}_k|} e^{-(\mathbf{F}_n-\mathbf{\mu}_k)^H\mathbf{\Sigma}_k^{-1}(\mathbf{F}_n-\mathbf{\mu}_k)}
\end{equation}
where $^H$ and $|.|$ denote conjugate transpose and determinant of a matrix, respectively. The model parameters $\mathbf{\theta}=\{\alpha_k,\mathbf{\mu}_k,\mathbf{\Sigma}_k\}_k$ are then estimated in the maximum likelihood (ML) sense via the expectation-maximization (EM) algorithm, which is well-known as an appropriate choice in this case, with the global log-likelihood defined as
\begin{equation}
\mathcal{L}_{ML}=\sum_{n=1}^N\log p(\mathbf{F}_n|\mathbf{\theta}).
\end{equation}

As a result, the parameters are iteratively updated via two EM steps as follow:
\begin{itemize}
\item E-step: compute the posterior probability that feature vector $\mathbf{F}_n$ is generated from the $k$-th GMM state
\begin{equation}
\gamma_{nk}=\frac{\alpha_k p(\mathbf{F}_n|\mathbf{\mu}_k,\mathbf{\Sigma}_k)}{\sum_{l=1}^K\alpha_l p(\mathbf{F}_n|\mathbf{\mu}_l,\mathbf{\Sigma}_l)}.
\label{eq:GMMEstep}
\end{equation}
\item M-step: update the parameters
\begin{align}
\alpha_k&=\frac{1}{N}\sum_{n=1}^N\gamma_{nk}\\
\mathbf{\mu}_k&=\frac{\sum_{n=1}^N\gamma_{nk}\mathbf{F}_n}{\sum_{n=1}^N\gamma_{nk}}\\
\mathbf{\Sigma}_k&=\frac{\sum_{n=1}^N\gamma_{nk}(\mathbf{F}_n-\mathbf{\mu}_k)(\mathbf{F}_n-\mathbf{\mu}_k)^H}{\sum_{n=1}^N\gamma_{nk}}.
\label{eq:GMMMstep}
\end{align}
\end{itemize}

With GMM, $N$ $d$-dimensional feature vectors $\mathbf{F}_n$ are characterized by $K$ set of GMM parameters $\{\alpha_k,\mathbf{\mu}_k,\mathbf{\Sigma}_k\}_{k=1,...,K}$ where $K$ is often very small compared to N. However, since GMM does not explicitly model the amplitude variation of sound sources, signals with different amplitude level but similar spectral shape may result in different estimated mean and covariance templates. \emph{To overcome this issue, another version of GMM called spectral Gaussian scaled mixture model (GSMM) could be considered instead. Though GSMM has been used in speech enhancement \cite{Hao-IEEETransASLP10} and audio source separation \cite{Benaroya-GMM06}, it has yet been applied in the context of fingerprinting}.

\subsection{HMM-based fingerprint}
\label{ssec:HMM}

HMM is a well-known model in many audio processing applications \cite{HMMtutorial}. When applied for audio fingerprinting, pdf of the observed feature vector $\mathbf{F}_n$ can be written as
\begin{align}
p(\mathbf{F}_n)=\sum_{q_1,q_2,...,q_d}&\pi_{q_1}b_{q_1}(F_{n,1})a_{q_1q_2}b_{q_2}(F_{n,2})\nonumber\\
&...a_{q_{d-1}q_d}b_{q_d}(F_{n,d})
\end{align}
where $\pi_{q_i}$ denotes the probability that $q_i$ is the initial state, $a_{q_iq_j}$ is state transition probability, and $b_{q_i}(F_{n,i})$ is pdf for a given state.

Given a sequence of observations $\mathbf{F}_n, n=1,...,N$ extracted from a labeled audio signal, the model parameters $\mathbf{\theta}=\{\pi_{q_i},a_{q_iq_j},b_{q_i}\}_{i,j}$ are learned via e.g., EM algorithm (detail formulation can be found in \cite{HMMtutorial}) and stored as a fingerprint. Cano et al. modeled MFCC feature vectors by HMM in their AudioDNA fingerprint system \cite{Cano-AES02}. In \cite{Cano-CME04} HMM-based fingerprint was shown to achieve a high compaction by exploiting structural redundancies on music and to be robust to distortions.

Note that when applying GMM or HMM for the fingerprint design, a captured signal at the user side is considered to be matched with an original signal fingerprinted by the model parameter $\mathbf{\theta}$ in the database if its corresponding feature vectors $\widehat{\mathbf{F}}_n$ are most likely generated by $\mathbf{\theta}$.

\subsection{NMF-based fingerprint}
\label{ssec:NMF}

NMF is well-known as an efficient decomposition technique which helps reducing
data dimension \cite{NMF_LeeandSeung}. It has been widely considered in audio and music processing, especially for audio source separation \cite{ontheflyaudio}\cite{interactive}. When applying in the context of audio fingerprinting, a $d \times N$ matrix of the feature vectors $\mathbf{V}=[\mathbf{F}_1,...,\mathbf{F}_N]$ is approximated by
\begin{equation}
\mathbf{V}=\mathbf{W}\mathbf{H}
\label{eq:NMF}
\end{equation}
where $\mathbf{W}$ and $\mathbf{H}$ are non-negative matrices of size $d\times Q$ and $Q \times N$, respectively, modeling the spectral characteristics of the signal and its temporal activation, and $Q$ is much smaller than $N$. The model parameters $\boldsymbol\theta=\{\mathbf{W}, \mathbf{H}\}$ can be estimated by minimizing the following cost function:
\begin{equation}
C(\boldsymbol\theta) =
\sum_{bn}d_{IS} \left([\mathbf{V}]_{b,n} | [\mathbf{WH}]_{b,n}\right),
\label{eq:crit1}
\end{equation}
where $d_{IS} \left(x|y\right)=\frac{x}{y}-\log\frac{x}{y}-1$ is Itakura-Saito (IS) divergence, and $[\mathbf{A}]_{b,n}$ denotes an entry of matrix $\mathbf{A}$ at $b$-th row and $n$-th column. The resulting multiplicative update (MU) rules for parameter estimation write \cite{Fevotte-NMF09}:
\begin{align}
  \mathbf{H} \leftarrow \mathbf{H} \odot
  \frac{\mathbf{W}^T\left(\left(\mathbf{W}\mathbf{H}\right)^{.-2}\odot\mathbf{V}\right)}
  {\mathbf{W}^T\left(\mathbf{W}\mathbf{H}\right)^{.-1}} \label{eq:Hupdate1} \\
  \mathbf{W} \leftarrow \mathbf{W} \odot
  \frac{\left(\left(\mathbf{W}\mathbf{H}\right)^{.-2}\odot\mathbf{V}\right)\mathbf{H}^T}
  {\left(\mathbf{W}\mathbf{H}\right)^{.-1}\mathbf{H}^T}
  \label{eq:Wupdate1}
\end{align}
where $\odot$ denotes the Hadamard entrywise product, $\mathbf{A}^{.p}$ being the matrix with entries $[\mathbf{A}]_{ij}^p$, and the division is entrywise.
Fingerprints are then generated compactly from the basis matrix $\mathbf{W}$, which has much smaller size compared to the feature matrix $\mathbf{V}$.

NMF was applied to the spectral subband energy matrix in \cite{Deng-ICCT11} and to the MFCC matrix in \cite{Chen-IET11}. The resulting fingerprint was shown to better identify audio clips than another decomposition technique namely singular value decomposition (SVD).


\section{Conclusion}
\label{sec:conclusion}

In this paper, we presented a review of the existing audio fingerprinting systems which have been developed by numerous researchers during the last decade for a range of practical applications. We described a variety of audio features and reviewed state-of-the-art approaches exploiting them for the fingerprint design. Furthermore, the use of statistical models and decomposition techniques to reduce the global statistical redundancy of feature vectors, and therefore to decrease fingerprint size, was also summarized. As a result, the combination of different presenting features and/or the deployment of a statistical feature model afterward are both applicable to obtain a robust and compact audio signature.

\bibliographystyle{IEEEtran}
\balance
\bibliography{AudioFP,references}

\end{document}